\documentstyle[art12,epsf]{article}

\catcode`\@=11

\def\refname{References}

\def\abstractname{Abstract}

\def\mathrm#1{{\rm #1}}

\newif\if@capcas \@capcasfalse
\newif\if@TwoColumn
\@TwoColumnfalse

\let\cchk@title\relax
\let\cchk@subtitle\relax
\let\cchk@openauthor\relax
\let\cchk@closeauthor\relax
\let\cchk@opencollab\relax
\let\cchk@closecollab\relax
\let\cchk@firstauthor\relax
\let\cchk@authorgroup\relax
\let\cchk@openaddress\relax
\let\cchk@closeaddress\relax
\let\cchk@addressgroup\relax
\let\cchk@oid\@gobble
\let\cchk@orf\@gobble
\let\cchk@abstract\relax
\let\cchk@openhist\relax
\let\cchk@received\relax
\let\cchk@revised\relax
\let\cchk@closehist\relax

\def\thebibliography{\section*{\refname}\@thebibliography}

\def\@thebibliography#1{\small
\list{\@biblabel{\arabic{enumiv}}}{\settowidth\labelwidth{\@biblabel{#1}}
    \leftmargin\labelwidth \advance\leftmargin\labelsep
    \itemsep 0pt plus 0.5pt minus 0.5pt
    \usecounter{enumiv}\let\p@enumiv\@empty
    \def\theenumiv{\arabic{enumiv}}}
  \def\newblock{\hskip 0.11em plus 0.33em minus -0.07em}
  \tolerance\@M \hyphenpenalty\@M \hbadness5000 \sfcode`\.=1000\relax}

\def\@biblabel#1{[#1]\hfill}

 \newtoks\t@glob@notes
\newtoks\t@loc@notes
\newcount\note@cnt
\newcounter{author}
\newcount\n@author
\def\n@author@{}
\newcounter{collab}
\newcount\n@collab
\def\n@collab@{}
\newcounter{address}

\newdimen\sv@mathsurround
\newcount\sv@hyphenpenalty

\newcount\prev@elem \prev@elem=0
\newcount\cur@elem  \cur@elem=0
\chardef\e@title=1
\chardef\e@subtitle=1
\chardef\e@author=2
\chardef\e@collab=3
\chardef\e@address=4

\newif\if@newelem
\newif\if@firstauthor
\newif\if@preface
\newif\if@articletype

\newbox\fm@box
\newdimen\fm@size
\newdimen\fm@margin
\newbox\t@abstract
\newbox\t@keyword

\let\report@elt\@gobble

\def\add@tok#1#2{\global#1\expandafter{\the#1#2}}

\def\add@xtok#1#2{\begingroup
  \no@harm
  \xdef\@act{\global\noexpand#1{\the#1#2}}\@act
\endgroup}

\def\beg@elem{\global\t@loc@notes={}\global\note@cnt\z@}

\def\@xnamedef#1{\expandafter\xdef\csname #1\endcsname}

\def\no@harm{
  \let\\=\relax  \let\rm\relax
  \let\ss=\relax \let\ae=\relax \let\oe=\relax
  \let\AE=\relax \let\OE=\relax
  \let\o=\relax  \let\O=\relax
  \let\i=\relax  \let\j=\relax
  \let\aa=\relax \let\AA=\relax
  \let\l=\relax  \let\L=\relax
  \let\d=\relax  \let\b=\relax \let\c=\relax
  \def\protect{\noexpand\protect\noexpand}}

\def\proc@elem#1#2{\begingroup
    \no@harm
    \let\thanksref\@gobble
    \@xnamedef{@#1}{#2}
  \endgroup
  \prev@elem=\cur@elem
  \cur@elem=\csname e@#1\endcsname
  \expandafter\elem@nothanksref#2\thanksref\relax}

\def\elem@nothanksref#1\thanksref{\futurelet\@peektok\elem@thanksref}

\def\elem@thanksref{\ifx\@peektok\relax
  \else \expandafter\elem@morethanksref \fi}

\def\elem@morethanksref#1{\add@thanksref{#1}\elem@nothanksref}

\def\add@thanksref#1{\global\advance\note@cnt\@ne
  \ifnum\note@cnt>\@ne \add@xtok\t@loc@notes{\note@sep}\fi
  \add@tok\t@loc@notes{\ref{#1}}}

\def\frontmatter{
  \global\t@glob@notes={}\global\c@author\z@
  \global\c@collab\z@ \global\c@address\z@
  \sv@mathsurround\mathsurround \m@th
  \global\n@author=0\n@author@\relax
  \global\n@collab=0\n@collab@\relax
  \global\advance\n@author\m@ne
  \global\advance\n@collab\m@ne
  \global\@firstauthortrue
  \global\@prefacefalse
  \global\@articletypefalse
  \open@fm \ignorespaces}

\def\endfrontmatter{\global\n@author=\c@author
  \global\n@collab=\c@collab \@writecount
  \global\@topnum\z@
  \thispagestyle{copyright}
  \if@preface \else
  \history@fmt
  \unvbox\t@abstract
  \cchk@abstract
  \fi
  \vskip 20pt
  \close@fm
  \output@glob@notes
  \global\@prefacefalse
  \global\leftskip\z@
  \global\@rightskip\z@
  \global\rightskip\@rightskip
  \global\mathsurround\sv@mathsurround
  \if@capcas \close@capcas \fi
  \setcounter{footnote}{0}
  \let\title\relax       \let\author\relax
  \let\collab\relax      \let\address\relax
  \let\frontmatter\relax \let\endfrontmatter\relax
  \let\@maketitle\relax  \let\@@maketitle\relax
  \@ifundefined{RIfM@}{}{\undo@AMS}\normal@text}

\let\maketitle\relax

\def\open@fm{\global\setbox\fm@box=\vbox\bgroup
  \hsize=\textwidth
  \sv@hyphenpenalty\hyphenpenalty
  \hyphenpenalty\@M}

\def\close@fm{\egroup
  \if@TwoColumn
    \emergencystretch=1pc
    \twocolumn[\unvbox\fm@box]\else \unvbox\fm@box
  \fi}

\def\output@glob@notes{\begingroup
  \ifx\@corresp\empty@data \else
    \corresp@note@fmt
    \footnotetext[1]{{\it Correspondence to\/}: \@corresp}\fi
  \the\t@glob@notes
  \endgroup}

\def\justify@off{\let\\=\@normalcr
  \leftskip\fm@margin \@rightskip\@flushglue \rightskip\@rightskip}
\def\justify@on{\let\\=\@normalcr
  \leftskip\fm@margin \@rightskip\z@ \rightskip\@rightskip}
\def\normal@text{\global\let\\=\@normalcr
  \global\leftskip\z@ \global\@rightskip\z@ \global\rightskip\@rightskip
  \global\parfillskip\@flushglue}

\def\@writecount{\write\@mainaux{\string\global
  \string\@namedef{n@author@}{\the\n@author}}
  \write\@mainaux{\string\global\string
  \@namedef{n@collab@}{\the\n@collab}}}

\def\title#1{
  \if@capcas \open@capcas \fi
  \beg@elem
  \title@note@fmt
  \add@tok\t@glob@notes
    {\title@note@fmt}
  \proc@elem{title}{#1}
  \cchk@title
  \def\title@notes{\the\t@loc@notes}
  \title@fmt{\@title}{\title@notes}
  \ignorespaces}

\def\subtitle#1{
  \beg@elem
  \proc@elem{subtitle}{#1}
  \cchk@subtitle
  \def\title@notes{\the\t@loc@notes}
  \subtitle@fmt{\@subtitle}{\title@notes}
  \ignorespaces}

\def\title@fmt#1#2{\vspace*{38pt}
  \bgroup \justify@off
    \Large\bf \noindent
    #1\,\hbox{$^{#2}$}\par
  \egroup}

\def\subtitle@fmt#1#2{\vspace*{5pt}
  \bgroup \justify@off
    \normalsize \noindent
    #1\,\hbox{$^{#2}$}\par
  \egroup}
\def\title@note@fmt{\def\thefootnote{\fnsymbol{footnote}}}

\def\corresp@note@fmt{\let\thefootnote\relax}

\def\author{\@ifnextchar[{\author@optarg}{\author@optarg[]}}

\def\author@optarg[#1]#2{\stepcounter{author}
  \beg@elem
  \@for\@tempa:=#1\do{\expandafter\add@thanksref\expandafter{\@tempa}}
  \report@elt{author}\proc@elem{author}{#2}
  \author@fmt{\the\c@author}{\the\t@loc@notes}{\@author}
  \cchk@openauthor
  \cchk@orf{#1}\cchk@closeauthor
  \ignorespaces}

\let\author@font\large

\def\author@fmt#1#2#3{\@newelemtrue
  \if@firstauthor \first@author \global\@firstauthorfalse \fi
  \ifnum\prev@elem=\e@author \global\@newelemfalse \fi
  \if@newelem \author@fmt@init \fi
  {\author@font #3\,$^{\mathrm{#2}}$}\nobreak}

\def\first@author{\author@note@fmt
  \add@tok\t@glob@notes
    {\author@note@fmt}
  \cchk@firstauthor}

\def\author@fmt@init{
  \cchk@authorgroup
  \par
  \vskip 25pt
  \justify@off
  \noindent}

\def\and{\ {\author@font and}\ }
\def\And{\par \vskip 10pt \noindent {\author@font and}\par
  \cur@elem=\e@address}

\def\collab{\@ifstar{\@tempswafalse}{\@tempswatrue}\collab@arg}

\def\collab@arg#1{\stepcounter{collab}
  \if@firstauthor \first@collab \global\@firstauthorfalse \fi
  \beg@elem
  \proc@elem{collab}{#1}
  \collab@fmt{\the\c@collab}{\the\t@loc@notes}{\@collab}
  \if@tempswa
    \cchk@opencollab\cchk@orf{#1}\cchk@closecollab\@tempswafalse
  \else\relax\fi
  \ignorespaces}

\def\collab@fmt#1#2#3{\@newelemtrue
  \ifnum\prev@elem=\e@collab \global\@newelemfalse \fi
  \if@newelem \collab@fmt@init \fi
  {\normalsize #3\,$^{\mathrm{#2}}$}}

\def\first@collab{
  \collab@note@fmt
  \add@tok\t@glob@notes
    {\collab@note@fmt}
  \cchk@firstauthor}

\def\collab@fmt@init{
  \cchk@authorgroup
  \par
  \vskip 1em
  \justify@off
  \noindent}

\def\author@note@fmt{\setcounter{footnote}{0}
  \def\thefootnote{\arabic{footnote}}}
\let\collab@note@fmt=\author@note@fmt

\def\address{\@ifstar{\address@star}
  {\@ifnextchar[{\address@optarg}{\address@noptarg}}}

\def\address@optarg[#1]#2{\refstepcounter{address}
  \beg@elem
  \report@elt{address}\proc@elem{address}{#2}
  \address@fmt{\the\c@address}{\the\t@loc@notes}{\@address}\label{#1}
  \cchk@openaddress
  \cchk@oid{#1}\cchk@closeaddress
  \ignorespaces}

\def\address@noptarg#1{\refstepcounter{address}
  \beg@elem
  \proc@elem{address}{#1}
  \address@fmt{\z@}{\the\t@loc@notes}{\@address}
  \cchk@openaddress
  \cchk@closeaddress
  \ignorespaces}

\def\address@star#1{
  \beg@elem
  \proc@elem{address}{#1}
  \address@fmt{\m@ne}{\the\t@loc@notes}{\@address}
  \cchk@openaddress
  \cchk@closeaddress
  \ignorespaces}

\def\theaddress{\alph{address}}

\def\address@fmt#1#2#3{\@newelemtrue
  \ifnum\prev@elem=\e@address \@newelemfalse \fi
  \if@newelem \address@fmt@init \fi
  \noindent \bgroup \small\it
  \ifnum#1=\z@
    #3\,$^{\mathrm{#2}}$\space
  \else
    \ifnum#1=\m@ne
      \@hangfrom{$^{\phantom{0}}$\space}
                {#3\,$^{\mathrm{#2}}$}
    \else
      \@hangfrom{$^{\mathrm{\theaddress}}$\space}
                {#3\,$^{\mathrm{#2}}$}
    \fi
  \fi
  \par \egroup}

\def\address@fmt@init{
  \cchk@addressgroup
  \par
  \vskip  6pt plus 2pt}

\def\abstract{\normal@text
  \hyphenpenalty\sv@hyphenpenalty
  \global\setbox\t@abstract=\vbox\bgroup
  \vskip 20pt                          
  \small\rm
  \hangindent=18pt \hangafter=1
  {\bf \abstractname:}
  \ignorespaces}
\def\endabstract{\par \egroup}

\def\journal#1{\gdef\@journal{#1}}

\def\history@fmt{\bgroup
  \cchk@openhist
  \relax
  \cchk@closehist
  \egroup}

\def\ps@plain{\let\@mkboth\@gobbletwo
 \def\@oddhead{}
 \def\@evenhead{}
 \def\@oddfoot{\hfil {\rm\thepage} \hfil}
 \let\@evenfoot\@oddfoot}

\def\@hexuple#1#2#3#4#5#6{\vtop{
  \hbox to \textwidth{\strut \rlap{#1} \hfil {#2} \hfil \llap{#3}}
  \hbox to \textwidth{\strut \rlap{#4} \hfil {#5} \hfil \llap{#6}}}}

\newbox\logo@box
\newlength{\logo@width}
\relax

\def\half{{\textstyle {1\over2}}}

\def\d{\,\mathrm{d}}
\def\e{\,\mathop{\mathrm{e}}\nolimits}

\def\int{\intop}

\@namedef{ack*}{\par\vskip 3.0ex plus 1.0ex minus 1.0ex}

\@namedef{endack*}{\par}

\def\ps@copyright{\ps@empty}
\def\@bibitem#1{\item\if@filesw \immediate\write\@auxout
       {\string\bibcite{#1}{\theenumiv}}\fi\ignorespaces}
\catcode`\@=12

\journal{Nucl.\ Phys.}

\def\tr{\mathop{\mathrm{Tr}}\nolimits}

\begin{document}
\begin{flushright}
      \vskip 1.0in
      PSU/TH/101\\
      hep-ph/9305309\\
      May 24, 1993
      \\
\end{flushright}
\begin{frontmatter}
\title{Measuring transversity densities in singly polarized
      hadron-hadron and lepton-hadron collisions }
\author{John C. Collins}
\author{Steve F. Heppelmann}
\address{Physics Department, Pennsylvania State University,
         University Park, PA 16802, U.S.A.}
\author{Glenn A. Ladinsky}
\address{Department of Physics and Astronomy,
         Michigan State University,
         East Lansing MI 48824, U.S.A.}
\begin{abstract}
       We show how the transverse polarization of a quark
       initiating a jet can be probed by the azimuthal
       distribution of two hadrons (of large $z$) in the jet.
       This permits a twist 2 asymmetry in hard processes when
       only one of the initial particles is polarized
       transversely. Applications to hadron-hadron and
       lepton-hadron scattering are discussed.
\end{abstract}
\end{frontmatter}

\section {Introduction}

A notorious result in perturbative QCD is that many transverse
spin asymmetries are zero at the twist-2 level.  While this
provides an interesting window \cite{QS,JJ,BB} onto twist-3
effects, it also means that the twist-2 part of the distribution
of transversely polarized quarks in a transversely polarized
hadron is rather difficult to measure.

In this paper, we suggest a new method for measuring transverse
spin effects {\em at the leading (twist-2) level}, by probing
suitable polarization sensitive properties in jet fragmentation.
Our technique jointly probes the distribution of transverse spin
(strictly, transversity) of quarks in a transversely polarized
hadron and the polarization dependence of the fragmentation of a
quark jet.  It can therefore be used in an experiment in
which only one of the incoming particles is polarized.  The
polarization dependence is in the azimuthal distribution of
low-mass pairs of pions about the jet axis.

One experiment in which these measurements could be made is
deeply inelastic scattering of unpolarized leptons on
transversely polarized hadrons.  The European Muon Collaboration
(EMC) has already demonstrated \cite{EMCrho}
that two particle correlations in the region
that we advocate are readily measurable.  Therefore, the
correlation should be accessible to a polarized version of that
experiment.  The proposed HERMES experiment at HERA should also
be able to do the measurement.

Another obvious experiment is in high $p_{\perp }$
particle production in
singly polarized hadron-hadron collisions. This would be ideal at
the experiment proposed by the RHIC Spin Collaboration
\cite{RSC} for the
RHIC collider at Brookhaven, if sufficient resolution could be
obtained.  A fixed target hadron-hadron experiment might be too
low in energy for the reliable application of perturbative QCD.

The normal spin asymmetries measured in singly polarized
collisions are higher twist \cite{KPR}, that is, they are
suppressed by
a power of the large virtuality in the hard scattering.  The
reason is that transverse spin asymmetries are associated
with off-diagonal terms in the spin density matrix of a
quark (when a helicity basis is used).  But in QCD, quark
helicity is conserved within a typical hard scattering, so
that there are no interference terms of the form:
\begin{equation}
    \langle \mathrm{helicity} +\mid \mathrm{final\ state}\rangle
    \langle \mathrm{final\ state}\mid \mathrm{helicity}-\rangle .
\end{equation}
This result is true to all orders of perturbation theory provided
that
\begin {itemize}
\item Quark masses are negligible on the scale of $p_{\perp }$.
\item All couplings are vector or axial vector.  (Thus both QCD
and electroweak physics with the photon, W and Z are covered.)
\item Final state polarizations are not measured.
\end{itemize}

Examples of this result are the higher twist nature of the
transverse spin asymmetries in high $p_{\perp }$ inclusive single
pion production in singly polarized hadron-hadron collisions
and in totally inclusive deeply inelastic scattering (the
structure function $G_{2}$ \cite{G2}).

To measure the transversity distribution at the twist-2 level in
a standard model process, we must probe the transversity of at
least {\em two\/} quarks that participate in the hard scattering.
In this paper, we propose that one of these quarks can be in
the final state:
a measurement can be made can
be made of the transverse spin state of a {\em final\/}
state quark, that is, a quark that initiates a jet.
Furthermore, this can be done by examining the correlation
of two hadrons of large $z$ and small relative transverse
momentum in the jet.  There are several advantages: The
experiment can be done in a singly polarized collision.
Thus one can go to a kinematic region where dominant process
is of valence quarks scattering on relatively small $x$
gluons, so that the cross section can be large.  The
subprocess asymmetry can be large, and we expect a
substantial asymmetry in the fragmentation.

Another possibility is to measure something like an intrinsic
$k_{\perp }$ distribution in fragmentation.  This has a twist-2
transverse spin asymmetry, and is discussed in a separate
paper \cite{trfr-sgl}.

Related earlier work can be found in \cite{Nacht,older}.  In
addition, similar ideas to ours have been proposed by
Efremov, Mankiewicz and T\"ornqvist \cite{EMT}.  They
concentrated on the case of longitudinal polarization, where
one has to measure a three particle correlation to get a
leading-twist asymmetry.

\section {Factorization}

\subsection {Unpolarized}

\begin{figure}
\begin{center}\leavevmode\epsfbox{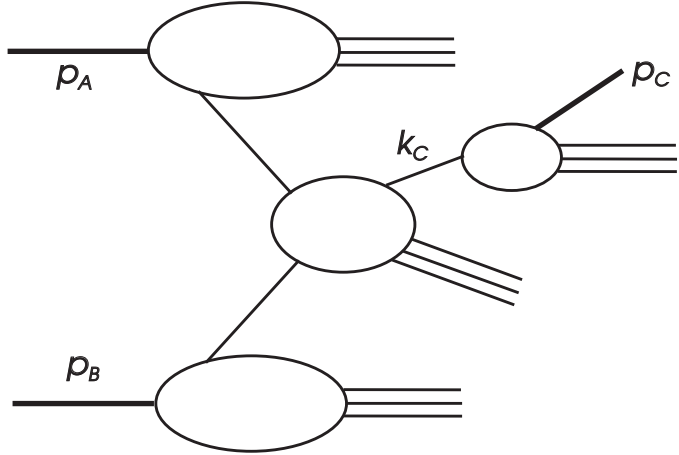}\end{center}
   \caption{Factorization theorem for single particle
      production in hadron-hadron collision.}
   \label{fig:fact}
\end{figure}

To define our notation and normalizations, let us recall the
factorization theorem for high $p_{\perp }$ inclusive single
particle production in hadron-hadron collisions $A+B\to C+X$.
Let the initial-state hadrons have momenta $p_{A}$ and $p_{B}$,
and let the observed hadron have momentum $p_{C}$. The
factorization theorem reads
\begin{equation}
  E_{C}\frac {\d\sigma }{\d^{3}{\bf p}_{C}} = \sum _{abc}
  \int  \d\xi _{A} \d\xi _{B} \frac {\d z}{z}\,
  f_{a/A}(\xi _{A})\, f_{b/B}(\xi _{B})\, |{\bf k}_{c}|
  \frac {\d\hat\sigma }{\d^{3}{\bf k}_{c}} \, D_{C/c}(z),
\label{eq:fact}
\end{equation}
which is illustrated in Fig.\ \ref{fig:fact}. The sum is over the various
flavors of parton (quarks, antiquarks and gluon) that can
participate in the hard scattering process, while $f_{a/A}$ and
$f_{b/B}$ are the parton densities for the initial hadrons, and
$D_{C/c}(z)$ is the fragmentation function.  The hard scattering
function $|{\bf k}_{c}|\d\hat\sigma /\d^{3}{\bf k}_{c}$ is for the
scattering $a+b\to c+X$ at
the parton level; it is a purely ultraviolet function, with all
mass singularities canceled, so that it can be calculated
perturbatively.  The variable $z$ represents the fractional
momentum of the measured hadron relative to its parent quark, so
that we set ${\bf k}_{c} = z {\bf p}_{C}$, when we use the
center-of-mass
frame of the hard scattering.  (Strictly speaking, we should use
a light-front definition.)
Corrections to Eq.\ (\ref{eq:fact}) are higher twist; that is, they are
suppressed by a power of the transverse momentum of $C$.

It can be checked that an interpretation of the
fragmentation function, with the normalizations indicated, is that
$zD_{C/c}(z) \d z$ is the number of hadrons of type $C$ in a parton
of type $c$ that have fractional momentum $z$ to $z+\d z$.  Because
of the factor $z$, it is common to define the fragmentation
function to be $d_{C/c}(z) \equiv  zD_{C/c}(z)$, rather than $D$.
However,
the behavior of $D$ under Lorentz transformations
is simpler, and this will be more convenient for us when we
define a function for the fragmentation into two observed
hadrons.

It might appear that we have neglected the possibility that the
hadron $C$ has transverse momentum relative to the parton $c$.
However, this is not so.  In accordance with the
derivation\cite{CSS}
of Eq.\ (\ref{eq:fact}), we have actually integrated over all small
values of this transverse momentum, while realizing that the
dependence of the hard scattering on small changes in the
transverse momentum is higher twist.  Large values of this
transverse momentum are correctly taken care of by the
higher order corrections to the hard scattering function.

\subsection {Polarized Case}

We now make two extensions to Eq.\ (\ref{eq:fact}).  The first is to take
account of polarization, and the second is to allow the observed
final-state hadron $C$ to be replaced by a two-particle state.

Polarization is easily taken account of.  First, whenever an
incoming hadron is polarized, the quarks and gluons entering the
hard scattering may themselves be polarized, and must be
equipped with a spin density matrix $\rho $ \cite{RS,AM,polfact}.
In the case of a massive on-shell
spin-$\half${} particle, like a proton, $\rho $ is completely
specified by the particle's Pauli-Lubanski spin vector.  But
in the case of a massless particle or of a parton, only the
density matrix itself may be used, since the Pauli-Lubanski
vector is singular in the massless limit.

The density matrix of the final-state parton $c$ is
determined by the initial-state polarization and by the
polarization dependence of the hard scattering cross section
$\hat\sigma $.  If the fragmentation depends on the spin of this
parton, then it will have some non-trivial analyzing power.
This can be represented by a matrix $\rho _{d}$, which we will call
the decay matrix.  It is the exact analog of a density
matrix in the initial state, except that it will be
normalized so that a fragmentation with no analyzing power
will have $\rho _{d}$ equal to the unit matrix.

In this paper, we will consider the case that only one of the
initial state hadrons is polarized, and that we make
a measurement sensitive to the spin of the final state parton
$c$.  Then Eq.\ (\ref{eq:fact}) must be replaced by
\begin{equation}
  E_{C}\frac {\d\sigma }{\d^{3}{\bf p}_{C}} = \sum _{abc}
  \int  \d\xi _{A} \d\xi _{B} \frac {\d z}{z}\,
  \rho _{\alpha \alpha '}f_{a/A}(\xi _{A})\, f_{b/B}(\xi _{B})\,
  H_{\alpha \alpha '\beta \beta '}(a+b\to c+X) \,
  \rho _{d,\beta \beta '}D_{C/c}(z),
\label{eq:trfact}
\end{equation}
where $H$ is the hard scattering cross section with the density
and decay matrices factored out.  Eq.\ (\ref{eq:trfact}) gives the cross
section for the process $A^{\uparrow}+B\to C+X$.  Its derivation is
a minor generalization of the derivation given in
\cite{polfact} for processes in which both of the
initial-state hadrons are polarized.

This equation represents the most general case.  Now
consider the case that hadron $A$ is transversely polarized.
Then the transversity of a quark in $A$ will be
proportional to the hadron's transversity.
But the gluons will be unpolarized,
because of conservation of angular momentum
about the axis of collision, as Artru and Mekhfi \cite{AM}
explained.
Now, transversity corresponds to the off-diagonal part of
the quark's density matrix.  So helicity conservation in the
hard scattering implies
that we need the final-state parton $c$ to be a quark, and
we need its decay matrix to be off-diagonal, if we are to
get a leading-twist asymmetry under reversal of the hadron's
transverse spin.\footnote{Note that a
   final-state gluon may acquire a polarization from the
   hard scattering, independently of the spin of the
   initial-state hadrons.  But this will produce no
   asymmetry under reversal of the hadron's polarization. }
Thus for the spin dependence, the sums over $a$ and $c$ in
Eq.\ (\ref{eq:trfact}) are only over flavors of quark and antiquark,
but the sum over $b$ is over gluons as well.

The standard measurement is for the observed final-state
hadron $C$ to be a pion.  Then there is certainly no
sensitivity of the fragmentation function to the
polarization of the quark, so that the decay matrix $\rho _{d}$ is
unity.  In this case, the asymmetry in the complete cross
section in Eq.\ (\ref{eq:trfact}) is zero.  The only asymmetry can come
from higher twist corrections, which are in fact
\cite{QS,BB} twist-3. This implies that, at large $Q$, the
asymmetry is roughly $M/Q$, where $Q$ is the scale of the
hard scattering, say the $p_{\perp }$ of the produced quark
(relative to the beam axis), and $M$
is some fixed hadronic mass scale.

But Eq.\ (\ref{eq:trfact}) applies to a general hadronic state $C$.
(The proof only requires that the state have an invariant
mass much less than $Q$.)  So to get a nonzero asymmetry at
the leading twist level, we merely need to find a state that
is polarization sensitive.  This is what we will do in the
next section.

\section {Polarization Sensitive Final States}

In this section, we discuss a number of candidates for
polarization-sensitive final states.

\subsection {Proton}

In principle we could choose the final-state hadron $C$ to be a
proton and measure its polarization.  This should give a
substantial asymmetry.  Indeed, it is known that in the
distribution of partons in a hadron, the flavor and spin states
of quarks at large $x$ reflect the corresponding quantities for
their parent hadron.  Similarly, in fragmentation, it is known
that the ratio of positive to negative pions at large $z$ in a
jet is substantially correlated with the charge of the initiating
parton \cite{EMCfrag}.  So it would be surprising if the contrary
were true for the spin state of hadrons at large
$z$ in a jet.

Unfortunately, the measurement of the spin of a proton of many
GeV of energy in a typical high-energy-physics detector is rather
hard.  (It is not impossible: one could conceive of measuring the
rescattering of the proton in the detector and using some
polarization sensitive process as a polarimeter.)  So we would do
better to try a different state for $C$.

\subsection {$\Lambda $}

The most obvious measurement \cite{AM}
is to pick out $\Lambda $s in the
final state, since their decay is self analyzing.  However,
they are produced rarely, and
in a simple quark model, the spin of the $\Lambda $ is carried by
its strange quark, which would substantially reduce the
asymmetry in the fragmentation.  This is by no means an
ironclad prediction, and is worth checking.

But the $\Lambda $ might result from the decay of a $\Sigma$, in
which case the above argument does not hold.

\subsection {$\Sigma^{0}$}

A better bet might be to measure the production of $\Sigma^{0}$,
with a measurement of the polarization of the $\Lambda $ in
its decay
serving to probe the spin of the $\Sigma$.  This seems a more
complicated, indirect method.

\subsection {$\rho $ meson}

One could imagine measuring production of $\rho $ mesons.  Now, the
density matrix of a massive spin-1 particle can be decomposed
into components of helicity 0, 1 and 2, and the
off-diagonal part of the density matrix of a quark
corresponds to a helicity flip of one unit. Therefore,
angular momentum conservation implies that only the
helicity 1 part of the density matrix of the $\rho $
gives the desired a transverse spin asymmetry in the
fragmentation. However, there is only one Lorentz invariant
in the amplitude of the decay $\rho \to \pi \pi $, so that
measurements
can only get at the {\em symmetric\/} part of the density
matrix, whereas the helicity 1 part of the density matrix is
antisymmetric.

It should be noted that the $\rho $ meson in fragmentation
appears over a rather larger background of continuum $\pi \pi $
states, and one could certainly have an interference effect.
This would appear as a special case of the inclusive two
pion measurement, which we will discuss next.

\subsection {Inclusive Two Pion}

Now let us consider the inclusive production of two pions, but
without requiring them to come from the decay of a $\rho $.  In
Eq.\ (\ref{eq:trfact}), we can use the following charge states of
nonidentical pions:
\begin{equation}
   C = \pi ^{+}\pi ^{0}\ \mathrm{or} \ \pi ^{-}\pi ^{0}
   \ \mathrm{or} \ \pi ^{+}\pi ^{-}.
\label{eq:2pichoice}
\end{equation}
The derivation of the factorization theorem does not care what
kind of a system the measured hadronic final state $C$ is,
provided only that its invariant mass is much less than the
natural scale of $p_{\perp }$ in the hard scattering.
Thus Eq.\ (\ref{eq:trfact})
continues to be valid. We may choose the invariant mass of the
two pion system to be in whatever range, say below a GeV or so,
that maximizes the asymmetry.  Presumably the appropriate range
will be where interference effects, e.g.,
$\rho $--$\omega $ $\rho $--continuum,
are important.

One can also measure the symmetric charge combinations,
e.g., $C= \pi ^{+}\pi ^{+}$, but the same argument as applied to the
$\rho $ will show that the asymmetry in the distribution of the
two pions must then be antisymmetric in the two pions, and
will therefore have a zero when the two pions have equal
values of their longitudinal momentum fractions.  This
suggests that the overall values of the asymmetry should be
less than when the two pions are unequal in charge.

Note also that the EMC \cite{EMCrho}
measured the distribution of two pion
systems in deeply inelastic scattering, as a function of the $z$
of the pair and of its invariant mass.  They observed a
substantial $\rho $ peak, but over a large continuum.  Thus the
unpolarized part of the fragmentation function to two pions,
$D_{\pi \pi /q}(z)$, is actually known.

To get at the polarization dependence, one must measure the plane
of the two pions.  For the larger invariant masses, one must
remember that the two pions can come out of two different jets,
and so there will be a correlation between the plane of the pions
and the plane of the scattering, and hence a non-uniform
azimuthal distribution.  This correlation will have some
tail down to low invariant masses.  What we are looking for is a
spin asymmetry in the azimuthal distribution of the plane.  We
will examine this in more detail in the next section.

\section {Fragmentation Functions}

Experimentally, the spin-dependence in Eq.\ (\ref{eq:trfact}) manifests
itself in the angular distribution of final-state particles.
To exhibit this, we write Eq.\ (\ref{eq:trfact}) in terms of
transverse spin vectors instead of density matrices.  We
define $s_{A\perp }$, $s_{a\perp }$, and $s_{c\perp }$ to be
respectively the
transverse spins of the initial hadron $A$, of the quark $a$
and of the quark $c$.  Our normalization is such that for
a fully polarized particle of helicity $\lambda $ and
transverse spin
$s_{\perp }$, we have $\lambda ^{2}-s_{\perp }^{2}=1$.

The decay matrix in Eq.\ (\ref{eq:trfact}) can be written in terms of
$\lambda _{d}$ and $d_{\perp }^{\mu }$,
which represent the analyzing power of the final state $C$
for the helicity and transverse spin of the quark initiating
the jet. The dependence on the spin of the quark $c$ is then
proportional to
$\lambda _{c}\lambda _{d}-s_{c\perp }\cdot d_{\perp }$,
of which we will only need
the $-s_{c\perp }\cdot d_{\perp }$ term in this paper.

In Feynman graph calculations of cross sections, the
spin-dependence of the hard scattering function can be
obtained by taking a trace of Feynman graphs times complex
conjugate graphs with projectors of the following form
\cite{BD}:
\begin{equation}
  \frac {1}{2}k\llap /
  (1+\lambda \gamma _{5}+\gamma _{5}s\llap /_{\perp }),
  \label{eq:diracsum}
\end{equation}
which represents the appropriate sum over quark wave
functions. The $1/2$ represents the spin average used for an
initial state, and is omitted for the final state.  More
modern calculations are done directly in terms of helicity
amplitudes \cite{GW}.

Then Eq.\ (\ref{eq:trfact}) can be rewritten as
\begin{eqnarray}
  E_{C}\frac {\d\sigma }{\d^{3}{\bf p}_{C}}
& = &
  \sum _{abc} \int  \d\xi _{A} \d\xi _{B} \frac {\d z}{z}\,
  f_{a/A}(\xi _{A})\, f_{b/B}(\xi _{B}) \,
\nonumber \\
&& \qquad
  \left[ H(a+b\to c+X) +
  A_{L}\lambda \lambda _{d} H_{L}(a+b\to c+X)
\right.
\nonumber \\ && \left. \qquad \qquad
  - A_{T}s_{A\perp }^{\mu }d_{\perp }^{\nu }
  H_{T\mu \nu }(a+b\to c+X) \right]
    \,D_{C/c}(z),
\label{eq:factfancy}
\end{eqnarray}
which is linear in the various spin vectors.
Here, $\lambda $ and $s_{A\perp }$ denote the
helicity and transverse spin
of hadron $A$.
We have used the
result that, in a parity invariant theory like QCD, the
helicity and transverse spin of quark $a$ are proportional to the
corresponding quantities for the parent hadron $A$;
the constants of proportionality are denoted by
$A_{L}$ and $A_{T}$.
Note that $A_{L}$ and
$A_{T}$ are functions of the parton variable $\xi _{A}$,
of the flavor $a$
of the initial quark, and of the Altarelli-Parisi scale $Q$ at
which the parton densities are computed.

We have used parity invariance together with helicity
conservation in Feynman graphs with massless quarks to show that
the hard-scattering cross section has the form of a
spin-independent term $H$ plus spin-transfer terms proportional
to $\lambda \lambda _{d}$ and $s_{\perp }\cdot d_{\perp }$.

For a given incoming parton state (particular values of $\xi _{A}$,
$\xi _{B}$, $a$, and $b$), the outgoing quark $c$ has helicity
\begin{equation}
  \lambda _{c} = A_{L} \lambda  \frac {H_{L}}{H},
\label{eq:finalhel}
\end{equation}
and transverse spin
\begin{equation}
   s_{c\perp \nu } =
     A_{T} s_{A\perp }^{\mu } \frac {H_{T\mu \nu }}{H}.
\label{eq:finaltrpol}
\end{equation}
The minus sign with the $H_{T\mu \nu }$ in Eq.\ (\ref{eq:factfancy})
arises because the
transverse spins are essentially Euclidean vectors.

\begin{figure}
\begin{center}\leavevmode\epsfbox{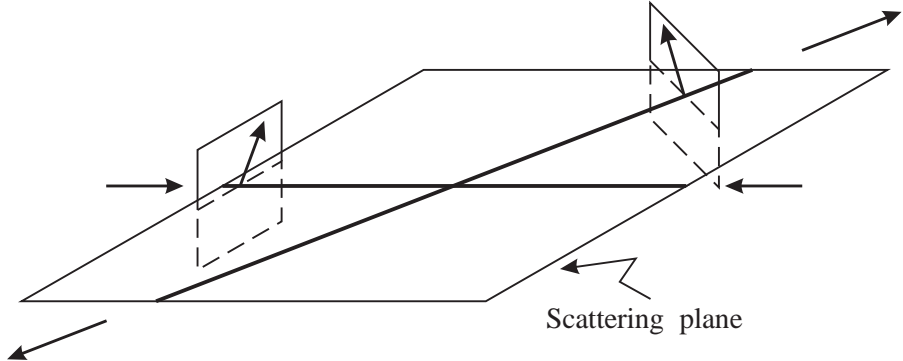}\end{center}
   \caption{Relation between spin vectors for initial and
      final state quarks.  }
   \label{fig:angles}
\end{figure}

An alternative form of the spin of the final-state quark is
\begin{equation}
     s_{c\perp }^{\mu } =
     T {s'}_{a\perp }^{\mu } = A_{T} T {s'}_{A\perp }^{\mu }.
\label{eq:spintran}
\end{equation}
Here, ${s'}_{a\perp }$ and ${s'}_{A\perp }$ are vectors
obtained by rotating
the initial quark and hadron spin vectors in the plane of
the hard scattering, as in Fig.\ \ref{fig:angles}.
We have used the parity invariance of
QCD, as applied to the hard scattering, to show that the
final-state quark spin is proportional to the rotated
initial-state quark spin, with a coefficient of
proportionality $T$ that we call the spin-transfer
coefficient.

\subsection{Operator Definitions}

To be able to derive Feynman rules for calculations with
parton distribution fragmentation functions, we need
operator definitions of these functions.  For the
unpolarized case, these were given in \cite{CS}, with a
motivation coming from light-front quantization.  The
definitions involve a light-like vector $n^{\mu }$ (a different
one for each of the parton densities and fragmentation
functions in Eq.\ (\ref{eq:trfact})).  All of these functions are
invariant under scaling of their $n^{\mu }$ vector, i.e., under
$n^{\mu }\to \lambda n^{\mu }$.

In the case of the distribution functions, this vector
together with the parent hadron's momentum $p^{\mu }$ defines the
axis of the scattering. If we let the light-like vector for
this case be $n_{a}^{\mu }$ then the parton density depends only
on the ratio $\xi \equiv k\cdot n_{a}/p\cdot n_{a}$, where $k^{\mu }$
is the momentum of the
(massless) parton entering the hard scattering.  This can be
rewritten in terms of the center-of-mass vector as
$\xi =k\cdot N/\sqrt {p\cdot N^{2}-p^{2}N^{2}}$.

Similarly, the fragmentation of a quark of momentum $k^{\mu }$
to a single hadron of momentum $p^{\mu }$ depends on one kinematic
variable $z=p\cdot n/k\cdot n$, where $n^{\mu }$ now represents the
appropriate light-like vector for the definition of the
fragmentation function.

Definitions for the polarized case were given in
\cite{RS,polfact,Man} for the distribution
functions.  Definitions of the polarized fragmentation
functions for the case that the transverse
momentum dependence is measured were given in
\cite{trfr-sgl}.  They are easily modified for our case.
The unpolarized fragmentation function is
\begin{equation}
   D_{C/c} \equiv
    \sum _{X} \int \frac {\d y^{-}}{12 (2\pi )^{3}}
        \e^{ik^{+}y^{-}}
     \tr \gamma ^{+}
     \langle 0| \, \psi _{c}(0,y^{-},0_{\perp }) \,|CX\rangle
     \langle CX|\,\bar \psi _{c}(0) \,| 0\rangle .
\label{eq:unpolfr}
\end{equation}
The factor of $1/12$ is the product of a factor $1/2$ which
occurs with all such definitions for fermions and a factor
$1/6$ for an average over the spin and colors states of the
quark. The sum over $X$ means that we sum
over all final states $|CX\rangle $
containing the chosen hadronic system $C$.  {\it Suitable
path-ordered exponentials of the gluon field along the line
$y^{{+}}=y_{\perp }=0$ must be inserted into this and the next
definition to make them gauge-invariant.}  We are using
light-front coordinates such that
$y^{\mu }y_{\mu }=2y^{+}y^{-}-y_{\perp }^{2}$; thus
the vector $n^{\mu }$ used above is $n^{\mu }=\delta _{-}^{\mu }$.

The definition (\ref{eq:unpolfr}) is easily generalized \cite{trfr-sgl}
to give the transverse
spin dependence $s_{c\perp }\cdot d_{\perp }D$ of the
fragmentation of a
polarized quark of transverse
spin $s_{c\perp }^{\mu }$:\footnote{The
   definition for the helicity dependence of the
   fragmentation will not concern us in this paper.}

\begin{equation}
   d_{\perp } D_{C/c} \equiv
  \sum _{X} \int   \frac {\d y^{-}}{12 (2\pi )^{3}}
    \qquad
    \e^{ik^{+}y^{-}}
    \tr \gamma ^{+}\gamma _{5}\, \gamma _{\perp }{\cdot }s_{\perp }
    \langle 0| \psi _{c}(0,y^{-},0_{\perp }) |CX\rangle
     \langle CX| \bar \psi _{c}(0) |0\rangle .
\label{eq:polfr}
\end{equation}
These definitions imply that the $k^{-}$
and $k_{\perp }$ of the incoming parton are integrated
over.  At first
sight Eq.\ (\ref{eq:polfr}) depends on two light-like vectors to define the
$+$ and $-$ axes.  In fact, it is invariant
under $s_{c\perp }^{\mu }\to s_{c\perp }^{\mu }+an^{\mu }$, so
that it actually only depends on one of the light-like vectors:
$s_{c\perp }^{\mu }$ has merely to be chosen
to satisfy $s_{c\perp }\cdot n=0$.

\subsection {Fragmentation to two hadrons}

When the observed state $C$ in the fragmentation function is
a single particle with no spin measured, it is easy to see
from Eq.\ (\ref{eq:polfr}) that there is no polarization dependence,
because from the vectors we have available we cannot
construct a non-zero scalar quantity linear in the quark's
transverse spin.\footnote{This requirement was evaded in
   \cite{trfr-sgl} by the measurement of the
   transverse momentum of $C$ relative to the quark: this
   effectively introduced another vector that played the role
   that in this paper is played by the relative transverse
   momentum within a two particle system for $C$.  Higher
   twist contributions to the cross section can also avoid
   this result, since more complicated Dirac matrix
   structures are then involved in the definition of the
   relevant fragmentation functions (compare \cite{JJ}).}

But we can construct such a quantity when $C$ is a
two-hadron state: it gives dependence on the azimuthal angle
between the plane of the two hadrons and the transverse spin
vector.
We let $p_{C1}^{\mu }$ and $p_{C2}^{\mu }$ be the momenta
of the two hadrons.
The unpolarized fragmentation
function, $D_{12/c}$ depends on three kinematic variables: the
invariant mass of the pair $m_{12}^{2}\equiv (p_{C1}+p_{C2})^{2}$,
and the fractional
momentum variables
$z_{1}\equiv p_{C1}\cdot n=
\sqrt {p_{C1}\cdot N^{2}-m_{C1}^{2}N^{2}}/k_{c}\cdot N$ and
$z_{2}\equiv p_{C2}\cdot n=
\sqrt {p_{C2}\cdot N^{2}-m_{C2}^{2}N^{2}}/k_{c}\cdot N$.
Here $k_{c}^{\mu }$ is the momentum of the
parton coming out of the hard scattering, and $N^{\mu }$ is the
rest vector of the hard scattering.

Define an axial vector:
\begin{equation}
   \Sigma_{\mu } \equiv
   \frac {\epsilon _{\mu \alpha \beta \gamma } n^{\alpha } p_{C1}^{\beta }
p_{C2}^{\gamma } }{(p_{C1}+p_{C2}^{})\cdot n \,\, m_{12} }.
\label{eq:axial}
\end{equation}
Then we get a scalar by contracting it with the quark spin:
$s_{c\perp }\cdot \Sigma$.
Hence we can write the spin-dependence of the fragmentation
in the form
\begin{equation}
   d_{\perp \mu }D(p_{1},p_{2},k_{c},N) =
   \Sigma_{\mu } A_{12/c}(z_{1},z_{2},m_{12}),
\label{eq:fragdep}
\end{equation}
where $A_{12/c}$ is a scalar function.  The factorization
theorem now becomes
\begin{eqnarray}
  E_{C_{1}}E_{C_{2}}
  \frac {\d\sigma }{\d^{3}{\bf p}_{C_{1}}\d^{3}{\bf p}_{C_{2}}}
  & = & \sum _{abc} \int  \d\xi _{A} \d\xi _{B} \frac {\d z}{z}\,
  f_{a/A}(\xi _{A})\, f_{b/B}(\xi _{B}) \,
\nonumber \\
  &&  \left[ H(a+b\to c+X) D(z_{1},z_{2},m_{12})
      \right.
\nonumber \\
   && \left.
           - A_{T}(\xi _{A}) A_{12/c}(z_{1},z_{2},m_{12})
           s_{A\perp }^{\mu }\Sigma^{\nu } H_{T\mu \nu }(a+b\to c+X)
       \right]
   .
\label{eq:fullfact}
\end{eqnarray}

No pseudoscalar can be constructed
with the helicity $\lambda _{c}$ and the vectors we
have available so
far.  Thus there is no leading-twist dependence of the
fragmentation of a longitudinally polarized quark to two
hadrons.\footnote{To
   get a helicity asymmetry, we need a third measured particle.
   Such asymmetries were considered by Nachtmann\cite{Nacht}, and
   more recently by Efremov et al.\cite{EMT}}
So we must set $\lambda _{d}$ in Eq.\ (\ref{eq:factfancy}) to zero.

The scalar product on which the fragmentation depends is
\begin{equation}
   -s_{c\perp }\cdot \Sigma =
    \frac {|k_{\perp }| \, |s_{\perp }|}{m_{12}} \cos \phi ,
    \label{eq:scalval}
\end{equation}
where $\phi $ is the angle between the transverse part of the
initial
hadron's spin vector and the normal to the plane of the two
detected final-state particles.  We have defined $k_{\perp }$ to be
the transverse momentum of one of the particles in $C$,
relative to the axis defined by $p_{C1}^{\mu }+p_{C2}^{\mu }$.
To be more precise, we
choose a $z$ axis along the spatial part of $p_{C1}+p_{C2}$.
We write
\begin{equation}
   {\bf p}_{C1} = (k_{\perp }, p_{C1}^{z}), \ \
   {\bf p}_{C2} = (-k_{\perp }, p_{C2}^{z}).
\label{eq:p12}
\end{equation}
Then
\begin{equation}
   \Sigma^{\mu } = (0,k_{\perp }^{y},-k_{\perp }^{x},0) / m_{12}.
\label{eq:sigval}
\end{equation}

In the unpolarized case, the plane of the two hadrons 1 and
2 is uniformly distributed in azimuth.  But Eq.\ (\ref{eq:scalval})
shows that in the case of transverse polarization there will
be a $\cos \phi $ asymmetry.

There are two sources of systematic error in this statement.  The
first is that the experimental apparatus is likely to be
asymmetric in this azimuthal angle about the axis
of the jet.  To cancel this, one should look for a change
in the angular distribution when the spin
of the initial-state hadron is reversed.

There is also theoretical error, since, when the
relative $k_{\perp }$ of
the final-state hadrons gets large enough, one should consider
that the hadrons
arise from the fragmentation of two different jets, as
in a $2\to 3$ parton process that is a higher order correction to
the lowest order $2\to 2$ process.  This will certainly give a
distribution that is not uniform in azimuth.  Again, one should
look to the asymmetry under reversal of the initial-state
spin to avoid these confounding effects.
The spin asymmetry will go away at large $k_{\perp }$,
since the 2 measured hadrons will then come from separate
jets, and for that situation we have seen that the spin
asymmetry is higher twist.

To check that the transverse spin asymmetry of the fragmentation
is indeed permitted by the symmetries of QCD, Collins and
Ladinsky \cite{CL} have calculated the fragmentation function of
a quark to two pions in a linear sigma model of quarks and pions.
This is of course by no means a perfect model, but it does have
the correct chiral symmetries.  Collins and Ladinsky do indeed
find a large spin asymmetry.  To get the asymmetry requires that
the quarks have a mass, as is obtained from chiral symmetry, and
that there be interference effects with non-trivial phases.
These interference effects occur between continuum two-pion
production and decay of the sigma resonance.  Analogous effects
are to be expected in more realistic models.

\section {Hard Scattering Calculations}

In this section we present the results of lowest order
calculations of the spin asymmetries for the hard scattering.

The calculation of the spin transfer coefficient is most easily
done by using a helicity decomposition of the scattering
amplitudes.  We let $A_{s_{a}s_{c}x}$ be the helicity amplitude,
with $s_{a}$
and $s_{c}$ denoting the helicities of quarks $a$ and $c$, and with
$x$ denoting the helicity state of all the other partons.  Then
the hard scattering coefficient in Eq.\ (\ref{eq:trfact}) is
\begin{equation}
    H_{\alpha \alpha ';\beta \beta '} \propto
    \sum _{x} A_{\alpha \beta x} A_{\alpha '\beta 'x}^{*}.
\label{eq:HfromA}
\end{equation}
Chirality conservation for massless quarks implies that
$H_{\alpha \alpha ';\beta \beta '}$ is nonzero only when:
\begin{itemize}
\item $\alpha =\alpha '$ and $\beta =\beta '$,
\end{itemize}
or when
\begin{itemize}
\item $a$ and $c$ are both quarks or both antiquarks of the same
flavor, $a=c$, and $\alpha =\beta $ and $\alpha '=\beta '$,
\item or $a$ and $c$ are the quark and antiquark of each other,
$a=-c$, and $\alpha =-\beta '$ and $\alpha '=-\beta $,
\end{itemize}
Thus the only nonzero terms are
$H_{++;++}$, $H_{--;--}$,
$H_{++;--}$, $H_{--;++}$,
$H_{+-;+-}$, and $H_{-+;-+}$.
The last two of these can only be nonzero if $a=\pm c$.
Parity invariance gives $H_{++;++}=H_{--;--}$,
$H_{+-;+-}=H_{-+;-+}$,
and $H_{++;--}=H_{--;++}$.

Then the spin transfer coefficient is
\begin{eqnarray}
    T &=& \frac {H_{+-;+-}}{H_{++;++}+H_{++;--}}
\nonumber \\
      &=& \frac {\sum _{x}A_{++x}A_{--x}^{*}}{\sum _{x}A_{++x}A_{++x}^{*}+\sum
_{x}A_{+-x}A_{+-x}^{*}}
\label{eq:TfromHA}
\end{eqnarray}
Formulae for the helicity amplitudes can conveniently be
found in the book by Gastmans and Wu \cite{GW}.

\subsection {Deeply Inelastic Scattering}

Exactly the same principles apply to deeply inelastic lepton
scattering as to hadron-hadron scattering, except that one of the
incoming particles is a lepton and hence can be treated as a
parton.  We let $p_{A}^{\mu }$ be the momentum
of the incoming hadron, and we
let $l^{\mu }$ and ${l'}^{\mu }$ be the
momenta of the incoming and outgoing leptons.

The calculation of the spin transfer coefficient to lowest order
was already done in \cite{DNN}.  We just use the tree graphs
for lepton-quark scattering with photon exchange.  The result is
\begin{eqnarray}
   T_{eqq} &=&
   \frac {4 (1 + \cos \theta )}{4 + (1 + \cos \theta )^{2}}
\nonumber \\
    &=& \frac {1-y}{1-y+\frac {1}{2}y^{2}} .
\label{eq:DIS}
\end{eqnarray}
Here $\theta $ is the scattering angle in the lepton-quark
center-of-mass, and $y$ is the
usual variable $p_{A}\cdot (l-l')/p_{A}\cdot l$,
which is $(E-E')/E$ in the target's rest
frame, and $(1-\cos\theta )/2$
in the lepton-quark center-of-mass.  The spin transfer
coefficient is 100\% at small scattering angles, and decreases to
zero for exactly backward scattering.  But even at $90^{\circ }$
($y=1/2$), it is 80\%.

\subsection {Hadron-Hadron Scattering}

There are several parton subprocesses $a+b\to c+X$, each of which
has its own spin transfer coefficient $T_{abc}$.  (Our notation
is that of the initial-state partons it is $a$ that is
polarized.)  As we have seen,
chirality conservation implies that $T_{abc}$ is zero
unless $a$ is a
flavor of (anti)quark and either $c=a$ or $c=-a$.  Here $-a$
denotes the antiparticle of $a$.

Thus at the tree level we need to
consider the following processes:
$i+g\to i+g$,
$i+i\to i+i$,
$i+\bar i\to i+\bar i$,
$i+\bar i\to \bar i+i$,
$i+j\to i+j$,
$i+\bar j\to i+\bar j$.
Here $j$ denotes
any flavor of quark that is not the same as $i$.  We took the
helicity amplitudes we needed from the book by Gastmans and
Wu \cite{GW}.

For the processes $qq'\to qq'$, with different quark flavors,
or for $qg\to qg$, we have
\begin{eqnarray}
   T_{qgq}= T_{qq'q} &=&
     \frac {4 (1 + \cos \theta )}{4 + (1 + \cos \theta )^{2}} =
      \frac {1-y}{1-y+\frac {1}{2}y^{2}} ,
   \label{eq:qqprimeq}
\\
   T_{qgg} = T_{qq'q'} &=& 0,
   \label{eq:qqprimeqprime}
\end{eqnarray}
the same as for electron-quark scattering.  Here, $\theta $ is the
angle between the incoming {\it polarized} quark and the
detected outgoing parton.  Note that in the
factorization formula each of these subprocesses will give two
contributions to the cross section, one with $c=q$, with the
spin transfer given by Eq.\ (\ref{eq:qqprimeq}), and one with $c=q'$ or
$c=g$, with zero spin-transfer.  We are assuming that of the
incoming quarks, it is the $q$ that is polarized.  The $q'$
may also be the antiquark of a flavor other $q$.

For identical quark flavors, we have
\begin{eqnarray}
   T_{qqq} &=& \frac {4 (1 + \cos \theta )^{2} (1+2 \cos \theta )}{(11 +
\cos^{2} \theta )\,(1 + 3\cos^{2} \theta )}
         = \frac {(1-y)^{2}(3-4y)}{(3-y+y^{2}) \, (1-3y+3y^{2})},
\label{eq:qqq}
\end{eqnarray}
and
\begin{eqnarray}
   T_{q\bar qq} &=&
   \frac {4 (1+\cos \theta ) \, (7-\cos\theta )}{35 + 8 \cos\theta  + 10
\cos^{2}\theta  - 8 \cos^{3}\theta  + 3\cos^{4}\theta }
   = \frac {(1-y) \, (3+y)}{3-2y+y^{2}-2y^{3}+3y^{4}},
\label{eq:qbarqq}
\\
   T_{q\bar q\bar q} &=&
   - \frac {4 (1+\cos \theta )^{2}}{35 - 8 \cos\theta  + 10 \cos^{2}\theta  +
8 \cos^{3}\theta  + 3\cos^{4}\theta }
\nonumber \\
   &=& - \frac {(1-y)^{2}}{3-6y+13y^{2}-10y^{3}+3y^{4}},
   \label{eq:qbarqbarq}
\end{eqnarray}
For unequal flavor annihilation, and for pure gluon initial
states, we get zero:
\begin{equation}
   T_{q\bar qq'} = T_{gga} = 0.
\end{equation}

Notice that, with one exception, all of the nonzero coefficients
equal unity at zero scattering angle, $y=0$.  The exception is
the coefficient, $ T_{q\bar q\bar q}$.

\section {Measurements}

We now summarize the measurements that can be done to probe the
polarized fragmentation function.  The distributions of quark
transversity in a hadron and the polarized fragmentation
functions are non-perturbative quantities for which we have no
detailed predictions.  But they should be universal: the
same in different processes.  Measurements of them will shed
light on the chiral properties of hadron wave functions and
of the long-distance part of the fragmentation.  In the
valence region for the distribution and fragmentation
functions, it is reasonable to conjecture from our
experience with other flavor quantum numbers that there are
large transverse spin correlations.

We are effectively trying measuring a spin transfer from the
initial state to the final state particle.  Suppose temporarily
that only one parton subprocess dominates, and that only a small
range of the parton variables is important.  Then the spin
asymmetry in the cross section is a product of the
asymmetries in its factors:
\begin{equation}
  A_{\sigma } = A_{f} T_{H} A_{D},   \label{eq:asym}
\end{equation}
where $A_{f}$ is the transverse spin asymmetry in the parton
density, $T_{H}$ is the spin transfer in the hard scattering
subprocess, and $A_{D}$ is the analyzing power of
the fragmentation.  The most
general case simply involves a weighted average of Eq.\ (\ref{eq:asym})
over the different subprocess
($\mathrm{quark} + \mathrm{gluon} \to
\mathrm{quark} + \mathrm{gluon}$,
$\mathrm{quark} + \mathrm{quark} \to
\mathrm{quark} + \mathrm{quark}$,
etc) and over the kinematic integrals.

The asymmetry for spin transfer in the hard subprocess can
be large, as shown by our calculations.
Moreover, in the valence region for the incoming
parton, we expect large polarization for the quarks.
Similarly, we might expect that at large $z$, say $z>0.5$
the spin of the final state hadron is highly correlated with
the spin of the quark initiating the jet, just as the flavor
is correlated. Thus, we might expect a large asymmetry in
the overall process.

\subsection {Deeply Inelastic Scattering}

Here one measures the distribution of two pions in a jet in
collisions of unpolarized electrons on transversely polarized
protons (or neutrons).

There should be a dependence of the azimuth of the plane of the
two pions that reverses sign when the spin of the incoming hadron
is reversed.  The azimuthal angle is about the jet axis.  If
the two pions are the leading pions in the jet, it should be
sufficient to define the jet axis by the sum of the pions'
momenta.
Maximum asymmetry will most likely occur under the
following conditions:
\begin{itemize}
\item When $x_{{\rm Bj}}$ is large, so that the quark entering
   the hard scattering will be highly polarized.
\item When the direction of the two pions is roughly that of the
   direction of the jet predicted by the parton model.  This
   avoids dilution by the zero asymmetry for pions in gluon jets.
\item When the fragmentation variable $z$ is large, so that
   the pions' quantum numbers follow those of the quark
   initiating the jet.  The two leading pions in the jet are the
   most obvious candidates.
\item When the charges of the pions are unequal.  Note that the
   asymmetry will reverse sign when the flavors of pions of equal
   longitudinal momentum are exchanged.
\end{itemize}
The asymmetry can also be measured for other flavor of hadron
pairs (e.g., involving kaons), for which there will be
corresponding polarized fragmentation functions.

The asymmetry has a kinematic zero when the invariant mass of the
pion pair is at threshold.  It will be higher twist when the
invariant mass of the pair is large, for then the pions will come
from different jets.

One test of the polarization dependence of the hard scattering
is in the dependence of the asymmetry on
the scattering angle $\theta $
in the lepton-quark frame, when $z$ and $x_{{\rm Bj}}$ are fixed.
This results from the $\theta $ dependence of the spin transfer
coefficient predicted by QCD in Eq.\ (\ref{eq:DIS}).

It is also possible \cite{DNN} to perform a measurement of the
azimuthal distribution of $\Lambda $ decays; this will give a
measurement of the polarization dependence of the
fragmentation of quarks to $\Lambda $s.

\subsection {Electron-Positron Annihilation}

In electron-positron annihilation, one can try measuring the
azimuthal correlations between a pair of pions one jet and a
pair in another jet \cite{EMT}.
Again, the pion pairs should be at large
fractional momentum: they should be leading pions to get the
biggest asymmetry.

The correlation will be proportional to the product of two
fragmentation asymmetries, and thus it will give a measurement of
the fragmentation asymmetry independently of the parton transversity
distribution in initial-state hadrons.  But the measurement will
require good control of the systematics, since there will be no
initial-state spin to reverse.

\subsection {Hadron-Hadron Scattering}

The measurement of the azimuthal distribution of pion pairs in
collisions of unpolarized hadrons on transversely polarized
nucleons is very similar to the case of deeply-inelastic
scattering.  The pion pair will need to be of low invariant mass
(say less than a GeV) but of large transverse momentum relative
to the incoming beams.  This
will ensure that the pair comes from a hard jet.
Again, the azimuthal angle is of the pion pair about the jet
axis.

A particularly favorable configuration is when the jet is
somewhat forward in the overall center-of-mass, so that the
hard collision would typically be of a
valence quark from the polarized
hadron and a small $x$ gluon from the unpolarized hadron.  The
valence quark would probably be highly polarized, and the
large number of small $x$ gluons will enhance the event rate.

As in the case of deeply inelastic scattering, one can
examine the dependence of the spin asymmetry on the angle
$\theta $ of the hard scattering, in the parton-parton
center-of-mass.  This would involve measuring opposite jets
associated with pion pairs so that the hard scattering can
be reconstructed.  (To a first approximation we will have
two-jet events.)

\section {Conclusions}

We have shown that the transverse spin of a quark initiating a
jet can be probed by the azimuthal dependence of pion pairs,
particularly leading pions, in the jet fragments.  This allows
leading twist spin asymmetries in singly polarized collisions.

Measurements can be made in deeply inelastic lepton-hadron
scattering and in hadron-hadron scattering.  They will give
jointly a transverse polarization asymmetry of quark
fragmentation and of the transversity distribution of quarks in a
transversely polarized hadron, and will test the spin transfer
coefficients predicted by QCD for the hard scattering.

\begin{ack}
This work was supported in part by the U.S. Department of Energy
under grant DE-FG02-90ER-40577, by the National Science
Foundation, and by the Texas National Laboratory Research
Commission.  We would like to thank Bob Carlitz,
Bob Jaffe and Xiang-Dong
Ji for many conversations and for collaborating with us in
the initial stages of this work.
We would also like to thank Frank
Paige, Jian-Wei Qiu, George Sterman and Mark Strikman for
useful discussions, and Xavier Artru for pointing out an error in
the calculation of $T_{q\bar q\bar q}$ in the original draft of this
paper.
\end{ack}

%
%

\end {document}